\documentclass[12pt]{article}

\usepackage{graphics,epsfig}
\usepackage{graphicx}
\begin{document}
\title{Heterogeneous believes,  segregation and  extremism in the making of public  opinions}

\author{Serge Galam
\\Laboratoire des Milieux D\'esordonn\'es et H\'et\'erog\`enes, \\
Tour 13, Case 86, 4 place Jussieu, 75252 Paris Cedex 05, }

\date{(galam@ccr.jussieu.fr)}
\maketitle

\begin{abstract}
The connection between contradictory public opinions, heterogeneous believes  and the emergence of democratic or dictatorial extremism  is studied extending our former  two state dynamic opinion model. Agents are attached to a social-cultural class. At each step they are distributed randomly in different groups within their respective class to evolve locally by  majority rule. In case of a tie the group adopts either one opinion with  respective probabilities $k$ and $(1-k)$. The value of $k$ accounts for the average of individual biases driven by the existence of heterogeneous believes within the corresponding class. It may vary from class to class. The process leads to extremism with a full polarization of each class along  either one opinions.  For homogeneous classes the  extremism can be along the initial minority making it dictatorial. At contrast heterogeneous classes exhibit a more balanced dynamics which results in a democratic extremism. Segregation among subclasses may produce a coexistence of opinions at the class level thus averting global extremism. The existence of contradictory public opinions in similar social-cultural neighborhoods is given a new light.
 \end{abstract}


\newpage

\section{Introduction}

In recent years the study of opinion dynamics has  become a main stream of research in Physics \cite{espagnol,huber,herrmann,slanina,rumor,stauffer,redner1,redner2,stauffer-meyer,mino,frank,weron,vicsek,sorin,deffuant,lemonde,chopard1}. Initiated long time ago \cite{strike, voting-old,mosco1} it is part of Sociophysics \cite{testimony}. 

Outside physics, research has concentrated on analyzing the complicated psycho-sociological
mechanisms involved in the process of opinion forming. In particular focusing on those
by which a huge majority of people  gives up to an initial minority view
\cite{1,2}. The main ingredient being, for instance in the case of a reform
proposal, that the prospect to loose definite advantages is much more energizing than
the corresponding gains which by nature are hypothetical.  

Such an approach is certainly realistic in view of the very active nature of minorities involved in a large spectrum of social situations. However we have shown \cite{mino,lemonde, chopard1} that even in the case of non active minorities, public opinion obeys some internal threshold dynamics which breaks its democratic character. Although each agent does have an opinion, they may find themselves into  local unstable doubting collective state while discussing with others in small groups. In such a case, we postulated that all group members adopt the same opinion, the one  which is consistent with the common believes of the group.
Examples of such common believes may be substantiated by saying like \textit{There is no smoke without a fire} or \textit{In case of a doubt, better do nothing}. 
Such a possibility of occurrence of local doubts results in  highly unbalanced conditions for the  competition of opinions within a given population even when each opinion has an identical convincing weight \cite{rumor,mino}.
 
In the present work we introduce social-cultural classes to include all agents which may meet together to discuss an eventual public issue. Each class is characterized by some common believes that result from the class average of all heterogeneous individual biases. Accordingly, in case of a tie in a local group, the resulting choice is  either one opinion with respective  probabilities $k$ and $(1-k)$ where $k$ accounts for the corresponding common belief of the social-cultural class. The value of $k$ is constant within each class with $0\leq k \leq 1$ and may vary from class to class.

Considering a class with a common belief $k$ and groups of size 4, $O$ denoting an opponent to the issue at stake and $S$ a supporter, our update rules writes,\\Ê

$SSSS$ and $OSSS \rightarrow SSSS$; $OOOO$ and $OOOS \rightarrow OOOO$,  \\ 

\begin{figure}
\begin{center}
\centerline{\epsfxsize=10cm\epsfbox{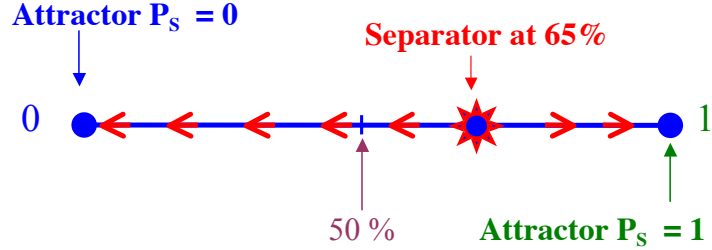}}
\caption{Opinion flow diagram with the two attractors at $0$ and $1$ and the separator  at a non symmetric value $65\%$. To survive a public debate one opinion must start at an initial support of more than fifteen percent while the other one needs an initial minority support. The associated extremism is dictatorial.
}
\end{center}
\label{flow}
\end{figure}

$ OOSS \rightarrow\Bigg \{ \begin{array}{c} OOOO $ with  probabiliy $k 
\\ \\SSSS$ with  probabiliy$(1-k) \end{array}$, \\\\\\
where all permutations are allowed. In our earlier works $k=1$ or $k=0$ except in \cite{chopard1} and in an application to cancerous tumor growth \cite{jan}. For $0\leq k\leq 1$ the opinion flow dynamics still converges towards a full opinion polarization but the separator location is now a function of $k$ and the group size distribution. It may vary from $0\%$ to $100 \%$ as shown in Fig. \ref{flow}.

The  flow is  fast and monotonous. When the initial majority makes the final opinion, the associated extremism is democratic, otherwise it is more or less dictatorial depending on the value of the initial minority which eventually spreads over the all class. Various extreme social situations can be thus be described.   In case of an homogeneous  high risk aversion sharing ($k=0$) it is found that a reform proposal may need an initial support of more than $90\%$ of the population to survive a public debate \cite{mino}. On the opposite, a shared high prejudice against some ethnic or religious groups ($k=1$) can make an initial false rumor shared by only a few percent of the population to spread over the whole population \cite{rumor}. Moreover very small fluctuations in the initial conditions may lead to opposite extremism. 

At contrast heterogeneous populations ($k=\frac{1}{2}$) are found to exhibit a more balanced dynamics. Extremism is turned democratic. At odd, segregation among subclasses leads opposite extremism within each subclass, which in turn stabilizes the associated overall class opinion at a permanent coexistence of both opinions  thus averting extremism.

Earlier version of  a threshold dynamical process can be  found in the study of  voting
in democratic  hierarchical systems \cite{voting-old,voting-recent}. There, groups of agents vote for
a  representative to the higher level using a local majority rule. 
Going up the hierarchy turns out to be exactly 
identical to an opinion forming process in terms of equations and dynamics. 
Instead of voting, agents update 
their opinions. The probability of electing  a representative at some hierarchy level
$n$ is equal to the proportion of  opinions sharing an opinion after $n$ updates 
\cite{chopard1,voting-old}. However within all these earlier studies the value of $k$ is always taken equal to one except in \cite{chopard1} where it was a function of the ration majority/minority within some finite size neighborhood.
The existence of threshold dynamics in social phenomena was  advocated
long time ago in  qualitative studies of some social phenomena \cite{schelling, grano}.

The rest of the paper is organized as follows. The model  is defined  in the next Section. It is then solved exactly within the simple case of groups of size three in Section 3. The counter intuitive case of groups of size 4 is studied in Section 4. In Section 5 small fluctuations are shown to sometimes lead to contradictory public opinions. It sheds a new light  on the fact that very similar areas in terms of their respective believes can hold opposite view, for instance about the feeling of safety.   Segregation effects are studied in Section 6 to find they can drive either democratic extremism or coexistence  of opinions thus avoiding global extremism. Including a size distribution for local groups is presented in Section 7. Last section contains some discussion.

\section{Setting the problem}

We start partitioning a given population among different social-cultural classes. Then we consider each class independently to study the dynamics of opinion forming among its  $N$ individual members facing some issue. It may be a
reform proposal, a behavior change like stopping smoking, a foreign policy
decision or the belief in some rumor.  The process is held separately within each class of agents. 

We discriminate between two levels in the process of formation of the global opinion, an external level and an internal one. The first one is the net result from the global information available to every one, the private information some persons may have and the influence of mass media. The second level concerns the internal dynamics driven by people
discussing freely among themselves. Both levels are interpenetrated but here we decoupled them to study specifically the laws governing the internal dynamics.  

Accordingly,  choosing a class of agents at a time $t$ prior to the public debate  the issue at stake is given a support by $N_s(t)$ individuals (denoted S) and an opposition from $N_o(t)$ agents
(denoted O).  Each person  is supposed to have an opinion with $N_s(t)+N_o(t)=N$. 
Associated individual  probabilities to be in favor or against the proposal at time
$t$ are,

\begin{equation}
p_{s,o}(t)\equiv \frac{N_{s,o}(t)}{N}  ,
\end{equation} 
with,
\begin{equation}
p_s(t)+p_o(t)=1 .
\end{equation}

From this initial configuration, people start discussing the project. However 
they don't meet all the time and all together at once. Gatherings are shaped 
by the geometry of social life within physical spaces like offices, houses, 
bars, restaurants and others. This geometry determines the number of people, which meet 
at a given place. Usually it is of the order of just a few. Groups may be larger 
but in these cases spontaneous splitting always occurs with people discussing in
smaller subgroups.  

To emphasize the  mechanism at work in the  dynamics which arises 
from local interactions no advantage is given to the 
minority with neither lobbying nor organized strategy. People discuss in small groups. To implement the psychological process of collective mind driven update, a local majority rule  is used within each group.  Moreover an identical individual persuasive power is assumed for both sides with the principle  ``one person - one argument''.  On this basis all members of a group adopt the  opinion which had the initial majority. In case there exists no majority, i.e. at a tie in a group of even size,  all members yet adopt the same opinion, but now either one with  respective probabilities $k$ and $(1-k)$ where $k$ is a function of the class.

\section{The intuitive case: the group of size three}

We first consider the case of update groups with the same size three. Accordingly to our 
local majority rule groups with either 3 S or 2 S ends up with 3 S. Otherwise
it is 3 O. The probability to find one supporter S after $n$ successive updates is,
\begin{equation}
p_s(t+n)=p_s(t+n-1)^3+3 p_s(t+n-1)^2 (1-p_s(t+n-1)) \ ,
\end{equation}
where $p_s(t+n-1)$ is the proportion of supporters S at a distance of $(n-1)$ updates
from the initial time $t$.

Eq. (3)  exhibits 3 fixed points  $p_{s,0}=0$, $p_{s,1}=1$. and $p_{c,3}=\frac{1}{2}$. First two corresponds to a total opinion polarization along respectively a total opposition to the issue
with zero supporters left and a total support. Both are attractors of the dynamics. The last point  with a perfectly balanced opinion splitting is  the separator of the dynamics as shown in Figs. (2) and (3).

\begin{figure}
\begin{center}
\centerline{\epsfxsize=10cm\epsfbox{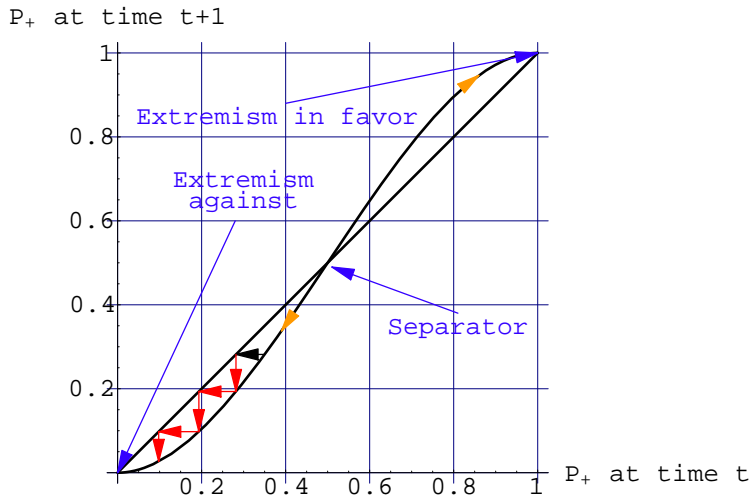}}
\caption{
Variation of the proportion $p_{s}(t+1)$ of supporters as function of $p_{s}(t)$ for groups of size $3$. Arrows show the direction of the flow for an initial support $p_s(t)<p_{c,3}=\frac{1}{2}$. The extremism is democratic.}
\end{center}
\end{figure}    

To reach the  attractor, the dynamics requires
a sufficient number of updates. In solid terms  each update
means some real time measured in numbers  of days whose evaluation is out the scope of the present work. An illustration is given starting  from $p_s(t)=0.45$. We get successively $p_s(t+1)=0.42$, 
$p_s(t+2)=0.39$, $p_s(t+3)=0.34$, $p_s(t+4)=0.26$, $p_s(t+5)=0.17$, 
$p_s(t+6)=0.08$ down to $p_s(t+7)=0.02$ and $p_s(t+8)=0.00$. Within 8 successive
updates, $45\%$ of the agents who were supporting the issue have shifted against it. The process has preserved and reinforced the initial majority making democratic the resulting extremism.The dynamics is perfectly symmetric with respect to both opinions as seen from Figs. (2) and (3). It is worth to stress that here the social-cultural character of the class is not activated since a local majority is always found.

\begin{figure}
\begin{center}
\centerline{\epsfxsize=10cm\epsfbox{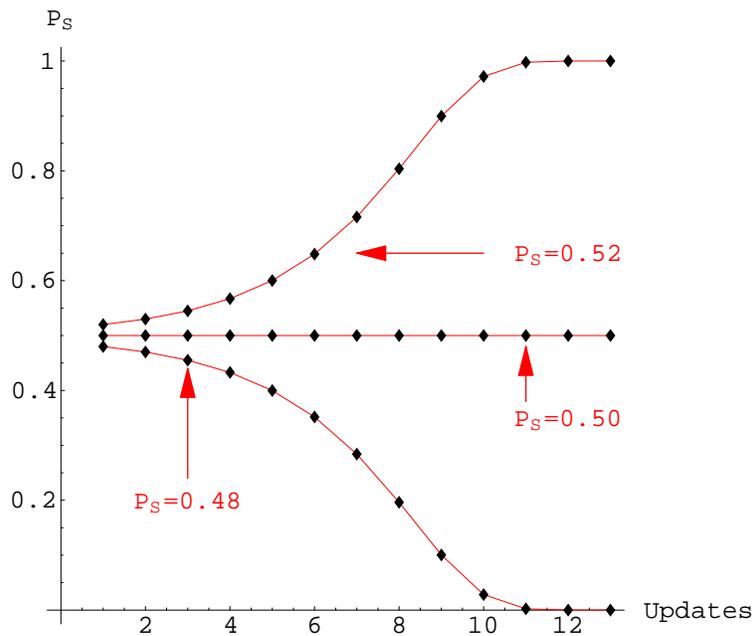}}
\caption{
Variation of $p_{s}(t)$ for groups of size 3 as function of repeated updates with three initial support $p_{s}(t)=0.48, 0.50, 0.52$. The resulting extremism is democratic since it is along the initial majority.}
\end{center}
\end{figure}

\section{The counter intuitive case: the group of size four}

It is only when dealing with even groups that the common believes driven bias can be analyzed.  There, the ``one person - one argument" rule allows for the possibility of a tie with no local majority. In such a case  participants are  in a non-decisional state. They are
doubting collectively, both opinions being supported by an equal number of arguments.  Here we evoke a common belief ``inertia principle"  to lift the doubt. We state that at a tie the group  eventually adopts the opinion O with a probability $k$ and the opinion S with the probability $(1-k)$ where $k$ accounts for the collective bias produced by the common believes of the group members. Some specific situations are considered below.

To illustrate the model we  consider  groups of size 4. The probability to find one
supporter S after 
$n$ successive updates becomes,
\begin{eqnarray}
p_s(t+n) \nonumber&= & p_s(t+n-1)^4+4 p_s(t+n-1)^3 \{1-p_s(t+n-1)\} \\
&  & 
+6 (1-k) p_s(t+n-1)^2 \{1-p_s(t+n-1)^2\ ,
\end{eqnarray}
where $p_s(t+n-1)$ is the proportion of supporters at a distance of $(n-1)$ updates from initial
time $t$. Last term includes the tie case contribution  (2S-2O) weighted with the probability $k$.

 \begin{figure}
\begin{center}
\centerline{\epsfxsize=10cm\epsfbox{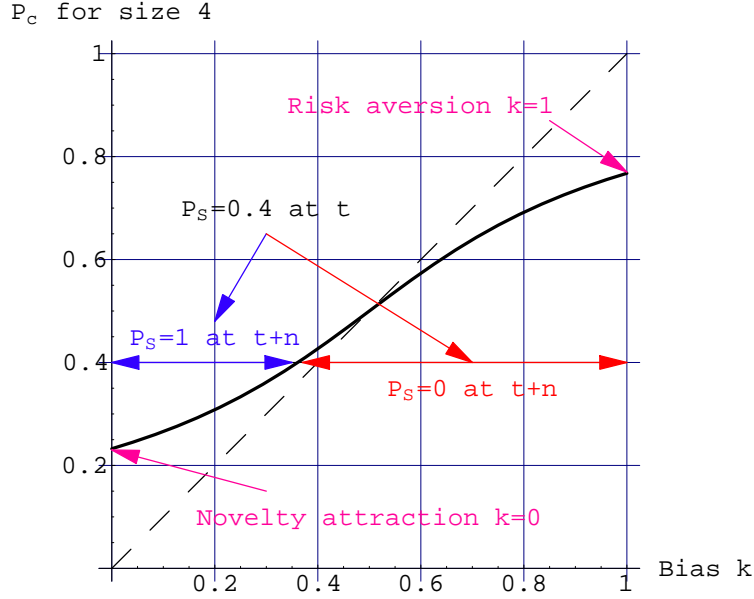}} 
 \caption{Variation of $p_{c,4}$ as function of $k$. For a collective risk aversion $(k=1)$, $p_{c,4}\approx 0.77$ while for  a collective novelty attraction $(k=0)$, $p_{c,4}\approx 0.23$. In case of no collective bias $(k=\frac{1}{2})$, $p_{c,4}=\frac{1}{2}$. An initial support of $p_{s}=0.40$ is shown to lead to a an extremism in favor ($p_S=1$) for the range of bias $0\leq k\leq 0.36$. At contrast the extremism is against ($p_S=0$) for the whole range $0.36\leq k\leq 1$.
}
\end{center}
\end{figure}    

From Eq. (4) both attractors $p_{s,0}=0$ and $p_{s,1}=1$ are recovered. However the unstable fixed point $p_{c,4}$
has now departed from the symmetric value  $\frac{1}{2}$ to the non-symmetric value,
\begin{equation}
p_{c,4}=\frac{(6k-5)+\sqrt{13-36k+36k^2}}{6(2k-1)} \ ,
\end{equation}
except at $k=\frac{1}{2}$ where $p_{c,4}=\frac{1}{2}$. Fig.  (4) shows the variation of $p_{c,4}$ as a function of $k$. The effect of the common believes of the class in the formation of the associated  public opinion is seen explicitly. For instance, an initial support of $p_{s}=0.40$ leads to a an extremism in favor the issue at stake  for the whole range of bias $0\leq k\leq 0.36$. At contrast the extremism is against it when $0.36\leq k\leq 1$.

For instance, when the issue relates to some reform proposal and the class shares a high risk aversion, a tie supports the Status Quo,  i.e.,  $k=1$ which yields $p_{c,4}\approx 0.77$.  The initial support for the reform to make the final public opinion has thus to start with more than $77\%$ as shown in Figs. (5) and (6)). When it does, the corresponding extremism is democratic.

\begin{figure}
\begin{center}
\centerline{
\epsfxsize=8cm\epsfbox{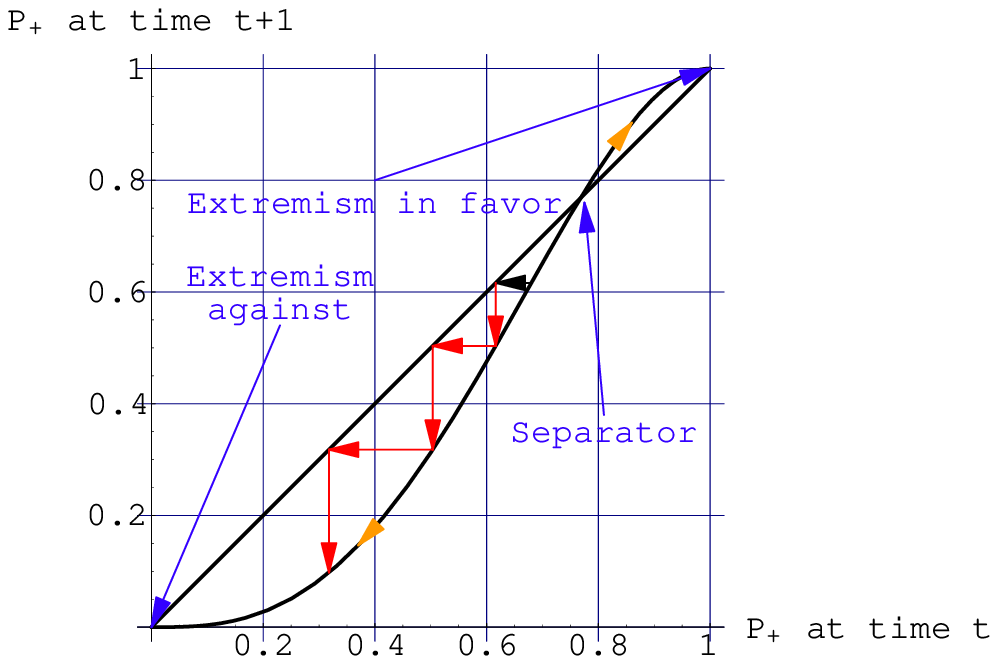}
\epsfxsize=8cm\epsfbox{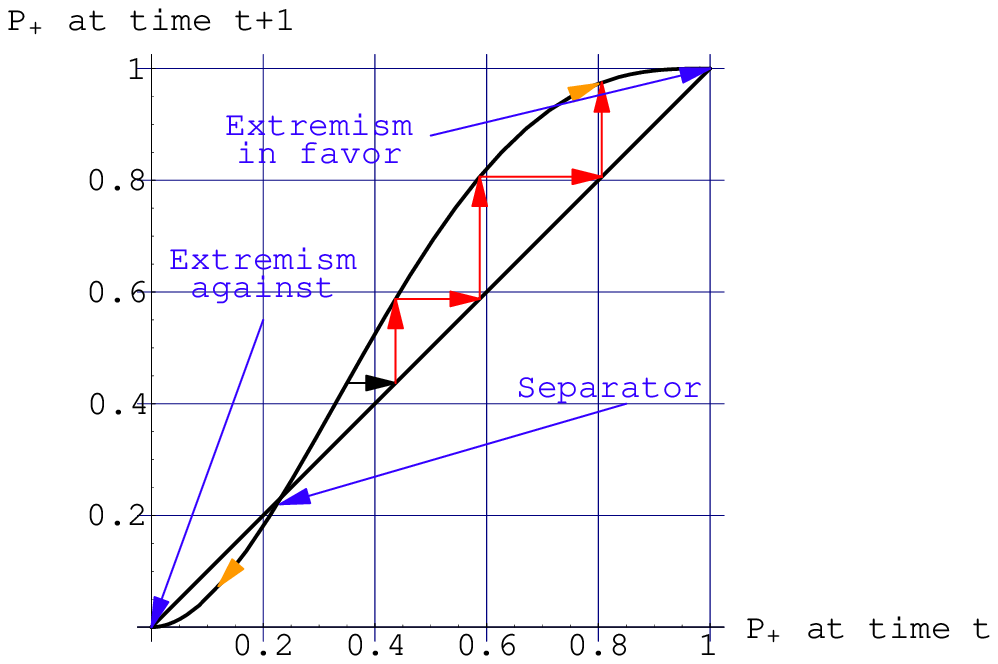}}
\caption{
Variation of $p_{s}(t+1)$ as function of $p_{s}(t)$ for groups of size 4. On the left side $k=1$ making $p_{c,3}\approx 0.77$. Arrows show the direction of the flow for an initial support $p_s(t)<0.77$. On the right side $k=0$ making $p_{c,3}\approx 0.23$. Arrows show the direction of the flow for an initial support $p_s(t)>0.23$. 
}
\end{center}
\end{figure}

On the other hand, an issue within a context where novelty is preferred drives a non-decisional state to adopt the opinion S making $k=0$ with in turn $p_{c,4}\approx 0.23$. An initial support of more than $23\%$ is now sufficient to invade the whole population. In case it happens, the resulting extremism is dictatorial since it is along an initial minority view. Associated flow dynamics are shown on the right side of  Figs. (5-6).

\begin{figure}
\begin{center}
\centerline{
\epsfxsize=7.0cm\epsfbox{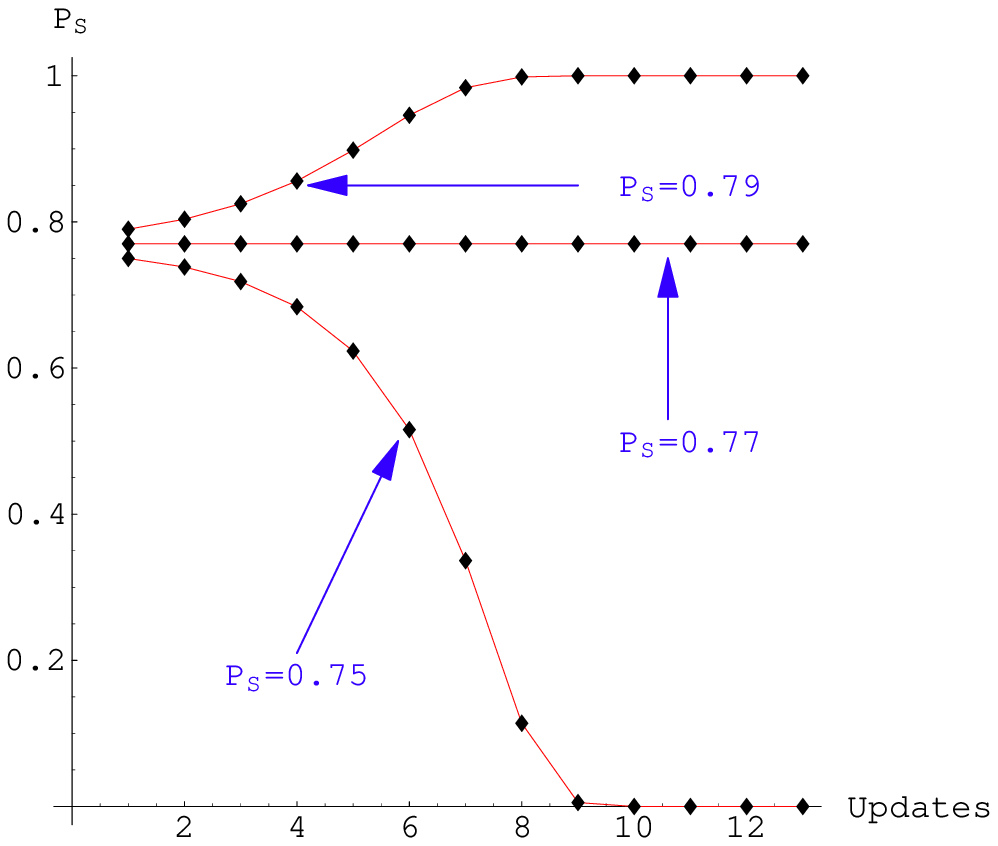}    
\epsfxsize=7.0cm \epsfbox{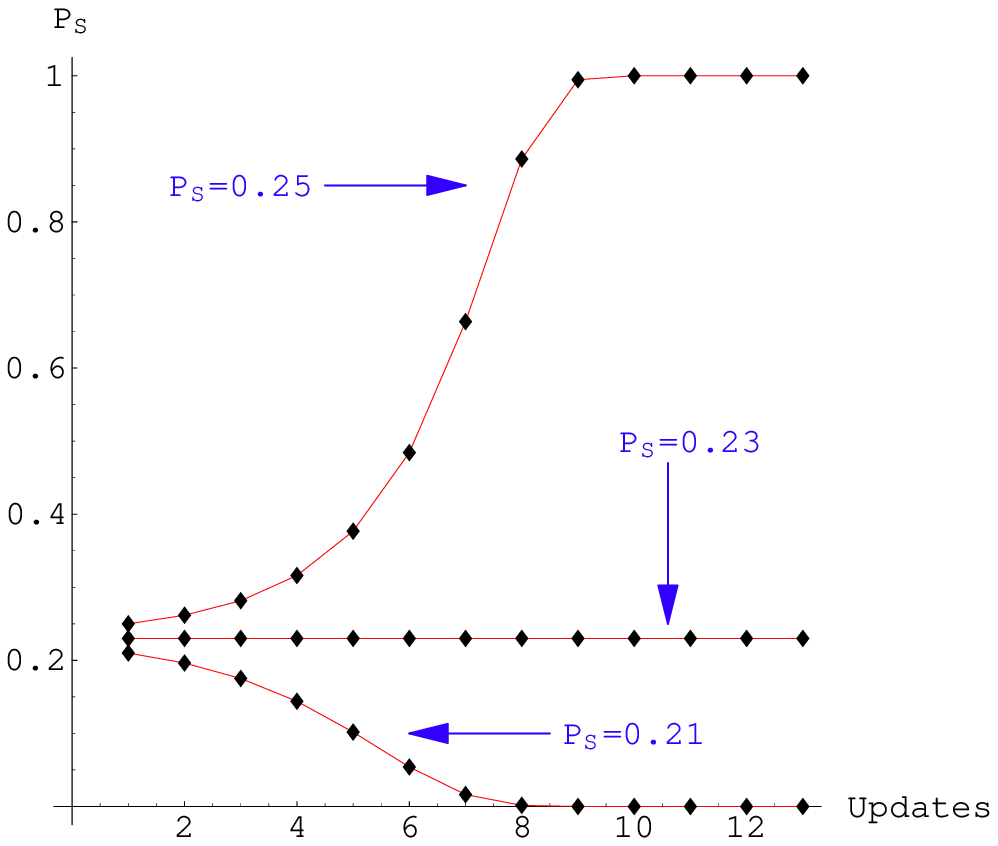}
}
\caption{
Variation of $p_{s}(t)$ for groups of side 4 as function of repeated updates. On the left size  $k=1$ with three initial support $p_{s}(t)=0.75, 0.77, 0.79$. To make the extremism, the initial support has to be larger than $77\%$. When it happens it is  democratic . 
On the right side $k=0$ with three initial support $p_{s}(t)=0.21, 0.23, 0.25$. The resulting extremism is dictatorial when it corresponds to an initial minority.}
\end{center}
\end{figure}    

In the case of groups of size 4 the number of updates to reach a full polarization is smaller than is the
case of size 3 as shown in Fig. (7) at $k=1$ and $k=0$. The number of required updates to have an extremism completed can be evaluated as, 

\begin{equation}
n\simeq \frac{1}{\ln[\lambda]}\ln[\frac{p_c-p_S}{p_c-p_+(t)}],
\end{equation}
where $\lambda$ is the first derivative of $p_s(t+1)$ with respect to $p_s(t)$ taken at $p_s(t)=p_c$. Moreover $p_S=0$ if $p_+(t)<p_c$   
while $p_S=1$ when $p_+(t)>p_c$. The number of 
updates being an integer, its value is obtained from Eq. (7) rounding to 
an integer. The number of updates diverges at $p_c$. The situation is symmetric with respect to  $k=0$ and $k=1$ with the divergence at respectively $p_c=0.23$ and $p_c=0.77$. It occurs at $p_c=0.50$  for $k=3$.

For instance, starting as above with groups of size 3 from $p_s(t)=0.45$ we get with $k=1$
the series  $p_s(t+1)=0.24$, 
$p_s(t+2)=0.05$ and $p_s(t+3)=0.00$. Within 3 successive
updates $45\%$  of support has shifted their support to  an opposition. Even an initial support above $50\%$ with $p_s(t)=0.70$ yields $p_s(t+1)=0.66$, $p_s(t+2)=0.57$, $p_s(t+3)=0.42$,
$p_s(t+4)=0.20$, $p_s(t+5)=0.03$, 
and $p_s(t+6)=0.00$. Only 6 updates are enough to have $70\%$ of supporters to shift their opinion.

\begin{figure}
\begin{center}
\centerline{\epsfxsize=10cm\epsfbox{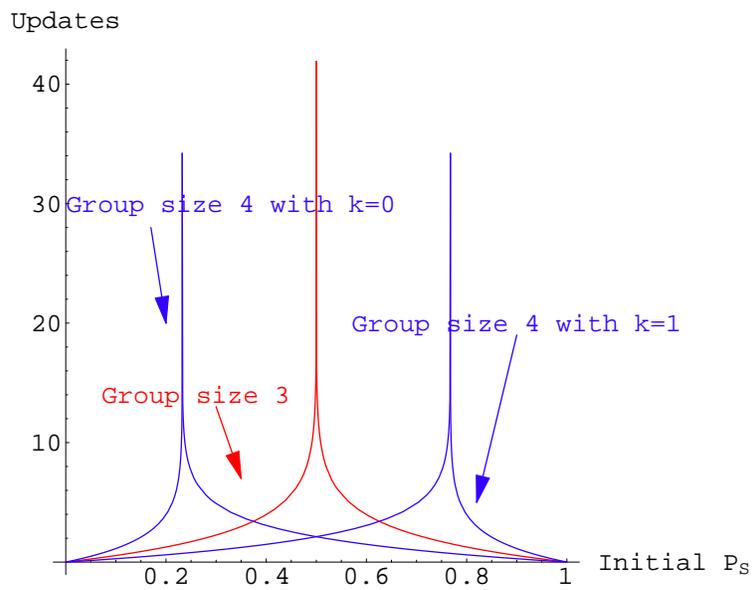}}
\caption{
Variation of the number of required updates to reach full extremism for groups of size 3 and  4 at $k=0$. 
}
\end{center}
\end{figure}    

\section{Small fluctuations and contradictory public opinions in similar areas}

We now discuss the effect of small differences of shared believes  in the making of public opinion of neighboring groups. For instance we consider a city area and its suburb as shown in Fig. (8). One class covers the city with  $k=0.49$ and another one the suburb with $=0.47$. Such a minor difference is not explicitly felt while crossing from one area to another. Both area are perceived as identical. However the study of the dynamics of opinion starting from the same initial conditions within each area shows that sometimes huge differences can be driven by either minor differences in the initial conditions. 

\begin{figure}
\begin{center}
\centerline{\epsfxsize=10cm\epsfbox{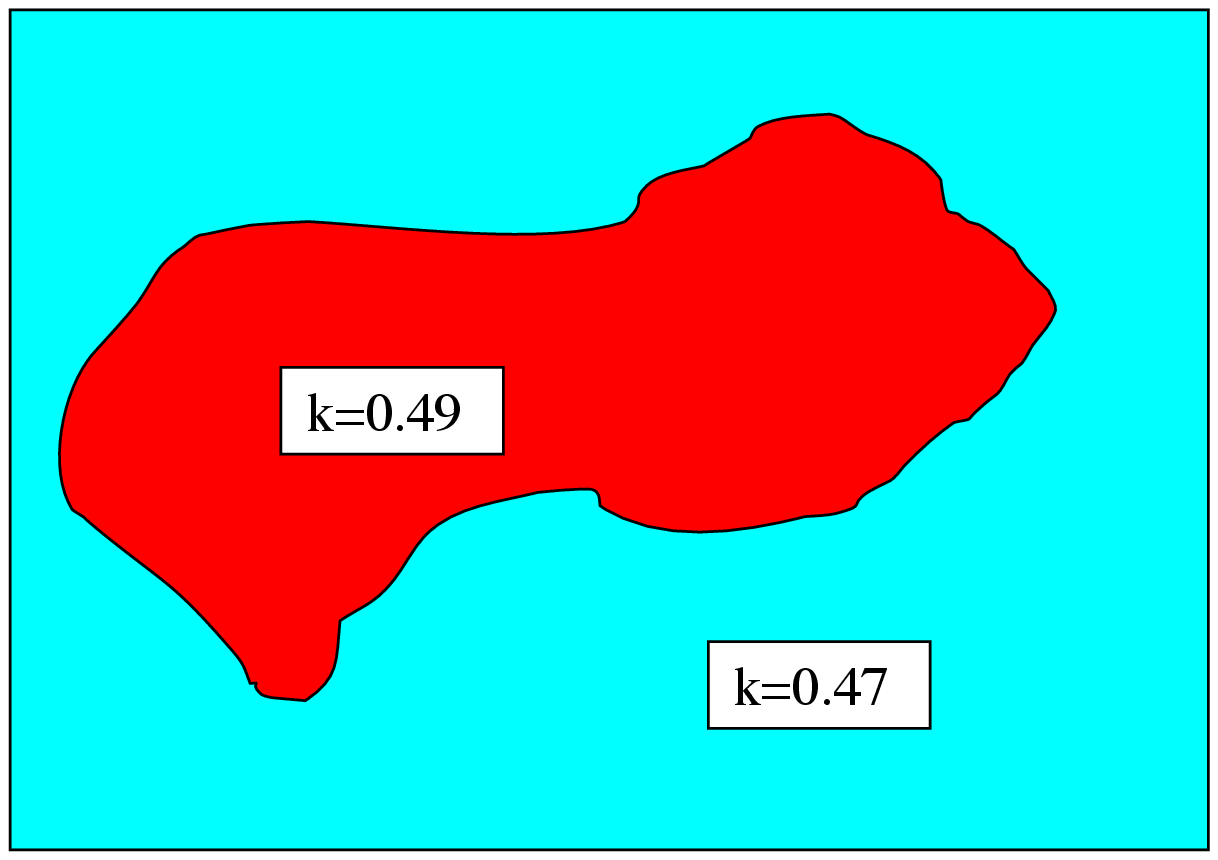}}
\caption{
A city with $k= 0.49$ and its surrounding suburb with $k=0.47$. 
}
\end{center}
\end{figure}

To illustrate our statement we consider two very similar initial conditions with an issue at stake having respectively $49\%$ and $51\%$ of support among both city and suburb populations. We then follow the dynamics of the corresponding public opinion flows. 

In the first case, $51\%$ of support to the issue results in both populations to a full support  to the issue making  both geographical areas identical as seen in Fig. (9). The process is completed within an estimate of ten updates. However a tiny decrease of $2\%$ in the initial support down at $49\%$ split the two neighboring similar areas. The city is now fully opposed to the issue while the suburb stays unchanged with a full support to it (see Fig. (9)).

\begin{figure}
\begin{center}
\centerline{\epsfxsize=10cm\epsfbox{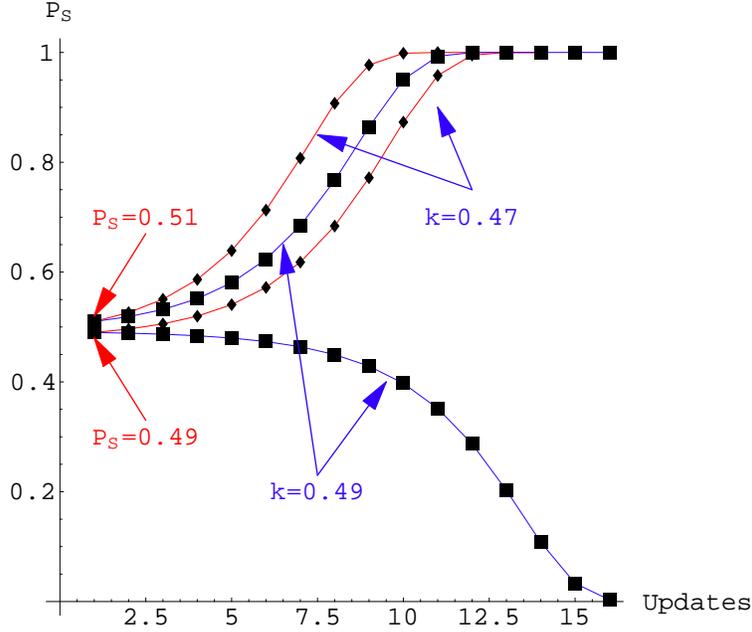}}
\caption{
Evolution of $p_s(t)$ as function of the updates for groups of size 4 with $k= 0.49$ and  $k=0.47$. Two initial supports are considered. For $p_s=0.51$ both $k= 0.49$ and  $k=0.47$ leads to $p_S=1$. But for $p_s=0.49$ only $k=0.47$ leads to $p_S=1$ while $k=0.49$ leads to $p_S=0$.
}
\end{center}
\end{figure}    

Above case may shed a new light on situations in which contradictory feelings or opinions are sustained  in areas which are nevertheless very similar like for instance the feeling of safety. It shows how an insignificant change in either the initial support or the bias driven by the common believes, may yield drastic differences in the outcome of  public opinion.  Fig. (10) shows the variations of the number of updates needed to reach extremism for several values of the bias $k$. Here too, the same initial support is shown to lead to totally different outcomes as function of $k$.
A value $p_S=0.30$ leads to  $p_S=0$ for both $k=0.50$ and $k=0.70$ while it yields $p_S=1$ for  $k=0.10$. For  $p_S=0.70$ all 3 cases lead to  $p_S=1$. 

\begin{figure}
\begin{center}
\centerline{\epsfxsize=10cm\epsfbox{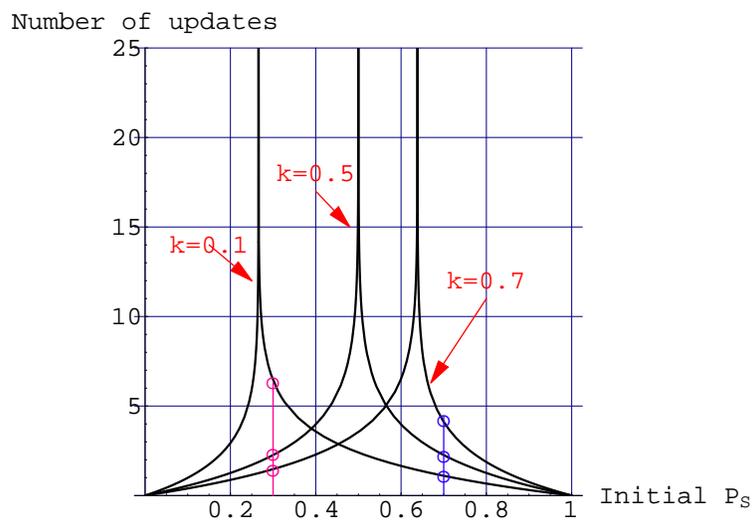}}
\caption{
Number of required updates to reach extremism for several values of the local bias with $k= 0.10$ $k=0.50$ and $k=0.70$. Associated values are shown for an initial support of respectively $30\%$ and  $70\%$
}
\end{center}
\end{figure}

\section{Segregation, democratic extremism and coexistence}

Up to now we have considered at a tie an average local bias $k$ which  results form a distribution of heterogeneous believes within a population. It means that all members of that population do mix together during the local group updates whatever is the individual respective believes. At this stage it is worth to note that different situations may arise in the distribution of the individual $k_i$. 

We discuss two cases for which either all $k_i$ are equal, i.e. an homogeneous population or they are all distributed among two extreme values for instance $0$ and $1$. There the existence of subclasses as a result of individual segregation may turn instrumental in producing drastic changes in the final global public opinion of the corresponding class. 

Consider first two different homogeneous classes  A and B in two different areas with respectively $k_i=0$ for all $i$ in A and $k_j=1$ for all $j$ in B.  From Eqs. (5-6) an initial support $p_s=0.25$ yields $p_S=1$ for A and $p_S=0$ for B as seen in Fig. (11). For A the extremism in support of the issue is dictatorial since along the initial minority $p_s=0.25$. At odd, in B the extremism is against the issue and democratic since along the initial majority of $1-p_s=0.75$ against it.

\begin{figure}
\begin{center}
\centerline{\epsfxsize=10cm\epsfbox{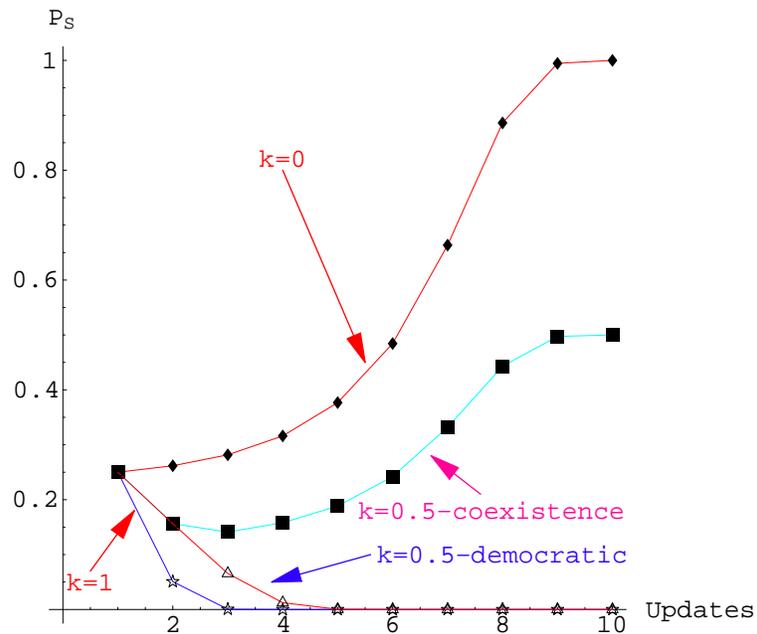}}
\caption{The dynamic of public opinions for two populations on different areas with respectively $k=0$ and $k=1$ at an initial support of  $p_s=0.25$ is shown on the upper and lower curves. In between the two populations are in the same area as subclasses of one unique class. The upper one has segregation and yields a coexistence of opinions. The lower one has mixing and reveals a democratic extremism. }
\end{center}
\end{figure}    

Now consider above classes A and B but as subclasses of the same class within one unique area. Two situations may occur as illustrated in Fig. (12) where A individuals are represented by circles and B ones by squares. They are white when in favor and black if against.  In the first situation (higher part of the Figure) people from each subclass do not mix together while updating their individual opinions. They are segregating each other yet sharing the same class within the same geographical area. As a result  two opposite extremism for each subclass is obtained as above. However the novelty with respect to distinct geographical areas is that here the resulting public opinion of the global class which does include the whole population is no longer exhibiting any extremism. The dynamics of segregated updates of opinions has produced a stable coexistence of both initial opinions with thus a balance collective stand. A poll over the bias would reveal the average value $k=1/2$.

In the second situation (lower part of the Figure) people from each subclass do mix together while updating their individual opinions. As a result, at a tie with mix individuals, A people adopt the opinion in favor while B people go on against. With two sub-populations with more or less the same large enough size this process is equivalent on average to have a bias $k=1/2$. The resulting extremism is democratic since it is along the initial global majority among the whole population. Mixing or segregation within the same situation may thus lead to drastically different public opinions as illustrated in Fig. (12).

\begin{figure}
\begin{center}
\centerline{\epsfxsize=10cm\epsfbox{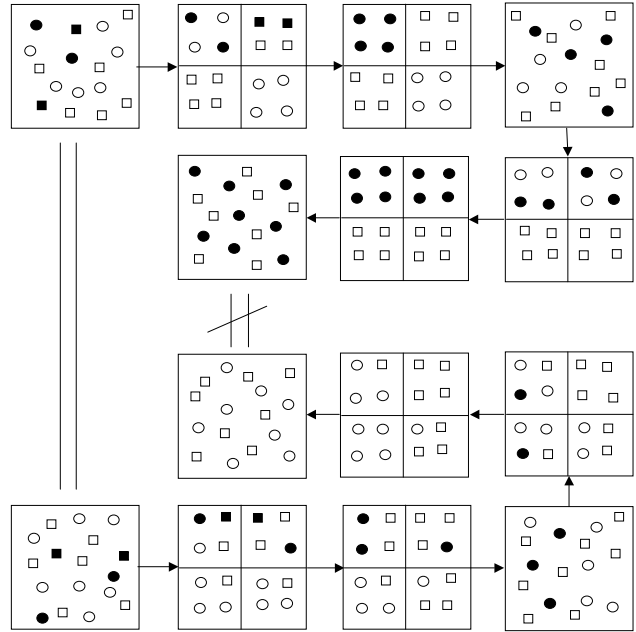}}
\caption{A population composed from two subclasses sharing opposite believes. The circles have $k=0$ while the squares have $k=1$. The initial proportion in favor (white) is identical for both of them at  $p_s=0.25$. Black color expressed an opinion against. In the upper series, peoples segregate while updating their opinions. The result is a perfect balance of the global public opinion with all circles against (blacks) and all squares in favor (whites). At contrast, in the lower series, circles and squares do mix together while updating. The result is a democratic extremism with all circles and squares in favor (white).}
\end{center}
\end{figure}

\section{Group size distribution}

Most of our analysis dealt with groups of size four. In real life people meet and discuss in several group sizes, from two up to five, six. Such a generalization was treated in  \cite{mino} in the case $k=1$. Along the same line we can define the general update Equation, 

\begin{eqnarray}
p_{s}(t+n)&=&\sum_{i=1}^L a_i \{\sum_{j=N[\frac{i}{2}+1]}^i C_j^i p_s(t+n-1)^j
p_o(t+n-1)^{(i-j)} \\
\nonumber&  & 
+(1-k_i)V(i)C_\frac{i}{2}^i p_s(t+n-1)^\frac{i}{2}
p_o(t+n-1)^\frac{i}{2}\},
\end{eqnarray}
where $C_j^i\equiv  \frac{i!}{(i-j)! j!}$, 
$N[\frac{i}{2}+1]\equiv $ Integer Part of  $(\frac{i}{2}+1)$, $p_{s}(t+n-1)$ is the
proportion of supporters after $(n-1)$ updates and
$V(i)\equiv {N[\frac{i}{2}]-N[\frac{i-1}{2} ]}$. It gives $V(i)=1$ for $i$ even and $V(i)=0$ for $i$ odd. We also have introduced the possibility of having the local bias $k_i$ to a function of the size $i$ of the  group.
Simultaneously, we have for the proportion of opponents, $p_{o}(t+n)=1-p_{s}(t+n)$.
The proportion of groups of size is defined by the probability distribution $a_i$  with the constraint $\sum_{i=1}^L a_i=1$ where $L$ is the largest group size and $i$ refers to the group size.

Clearly an infinite number of size distribution $\{a_i\}$ is possible. Also various value of $k$ can be considered depending on the size group as well as some agent dependence though the mixing of difference sub-classes. 
However the existence of local collective non-decisional state  monitored by the occurrence of local ties in groups
of even size  will always occur. Such a feature whatever its amplitude is does produce an asymmetry in the  polarization dynamics towards either one of the two competing opinions thus preserving the main result of the simple version of the model presented in this paper.

\section{Conclusion}
  
To conclude, we have presented a simple model which is able to reproduce some  complexity of the social reality. It suggests that  the direction of the
inherent polarization effect in the formation of a public opinion driven by a democratic debate
is biased from the existence of common believes within a population. Homogeneous versus heterogeneous situations were shown to result in different qualitative outcomes.
  
At this stage we did not address the difficult question on how to remedy this reversal opinion phenomenon with the natural establishment of dictatorial extremism. The first hint could be in  avoiding  the activation of common general background in the social representation of reality.
However direct and immediate votes could be also rather misleading. Holding an immediate vote without a debate as soon as a new issue arises has other drawbacks. At this stage, the collaboration with psycho sociologists as well as political scientists would be welcome.

In addition, in real life not every person is open mind and changes
opinion. Therefore it would be interesting to introduce stubborn agents in the model. The model may generalize to a large spectrum of social, 
economical and political phenomena that involve propagation effects. In 
particular it could shed a new light on both processes of fear propagation 
and rumors spreading.

\end{document}